# UNDERSTANDING THE COULOMB EXPLOSION THROUGH NEUTRALIZATION DYNAMICS IN THE PROBLEM OF AUGER-DESTRUCTION

[1*]Nigora N. Turaeva, [2]Boris L. Oksengendler, [3]Farida Iskandarova

[1]Department of Natural Sciences and Mathematics, Webster University, Saint Louis, USA 63119

[2]Institute of Materials Science of the Academy of Sciences of the Republic of Uzbekistan Tashkent, Uzbekistan 111310

[3]The Center for the Development of Nanotechnology, National University of Uzbekistan named after Mirzo Ulugbek, Tashkent, Uzbekistan 100174

*nigoraturaeva82@webster.edu

**Abstract**

The Coulomb explosion is a process that occurs following the formation of multiple charges during Auger cascades, leading to the destruction of solid-state and molecular structures. For unstable multi-charged ions produced at Auger cascades, the destruction cross-section is directly related to the probability of Coulomb explosion, which depends on characteristic times of ion dissociation and electron neutralization. This study demonstrates that the atomic dissociation time decreases with charge according to the relationship $1/Z^{1/2}$. An expression for the neutralization time was obtained within the model, in which the probability of Coulomb explosion is considered as a competition of two processes, ion dissociation and electron neutralization. By using approximation of the effective mass, it was shown that the neutralization time is a function of the effective mass and the width of the valence band of solid states.

The Auger cascade is a non-radiative process, in which a vacancy in an atom's inner electron shell—created by orbital electron capture, K-capture, or at absorption of a high energy photon—gets filled by an electron from a higher energy shell [1-3]. This triggers a series of sequential electronic transitions in the inner and outer shells, resulting in the emission of low-energy electrons rather than X-rays. A small molecule undergoes decay due to K-capture by the nucleus of one of its constituent atoms, which is a result of the Auger cascade [ 4]. Selective irradiation of specific atoms in DNA through inner-shell ionization, followed by an Auger cascade, leads to molecular degradation of DNA [5]. Biological studies of the Auger effect in DNA focused on constituent atoms like oxygen, nitrogen, carbon, and phosphorus, as well as externally administered bromine, iodine, and platinum [5,6]. The studies showed that the Auger cascade, occurring within a very narrow energy range near the K-shell absorption edge, significantly increased DNA damage, resulting in single-stranded breaks (SSBs) and double-stranded breaks (DSBs). It is also important to note that in all cases involving constituent atoms, positively charged ions were desorbed from the sample surface due to molecular degradation caused by irradiation [7]. Additionally, the significant role of Auger cascades in defect formation due to ionizing radiation has been confirmed in both bulk [8-13] and surface of crystals [14].

The Dexter-Varley paradox [11,12] addresses the potential for defect formation in crystals via Auger cascades. Generally, the unstable multiply charged state resulting from the Auger cascade decays through two competing channels: either through Coulomb explosion of ions leading to defect formation or through the neutralization of positive charges by valence electrons, which does not result in defects. Varley argued that in an ionic lattice subjected to ionizing radiation the defect is formed via Coulomb explosion in $\tau_+ \approx 5 \times 10^{-14} sec$. Dexter calculated the neutralization time for the positively charged ion in the ionic crystals based on the principle of uncertainties, $\Delta E_v \cdot \Delta t \geq \frac{\hbar}{2}$, and found that $\tau_e \approx 10^{-15} - 10^{-16} sec$, which was much less than the defect formation time, obtained by Varley. Therefore, Dexter suggested that the defect could not be formed in ionic crystals. This paradox highlights the complexities of competing processes in solid-state physics, particularly in the context of ionization and defect formation mechanisms. To resolve this paradox, by considering a quantum nature of the neutralization process, a simple approximation for the probability of defect formation was inductively introduced, that is $P = \exp(-\tau_+/\langle\tau_e\rangle)$, where $\langle\tau_e\rangle$ is the mean neutralization time [9]. This approach has helped explain numerous experiments on subthreshold defect formation by Auger cascades in metals, semiconductors, and insulators, effectively resolving the Dexter-Varley paradox [9]. We can note here that the experimental values of $\tau_e$ vary in a wide range: for metals $\langle\tau_e\rangle \approx 10^{-16} s$, for semiconductors $\langle\tau_e\rangle \approx 5 \cdot 10^{-15} s$; for insulators $\langle\tau_e\rangle \approx 10^{-13} s$. Thus $P_{metal} \ll P_{semicond} < P_{insult}$ [10].

According to the model [10], the overall process of destruction/defect formation by the Auger cascade consists of three sequential stages: the creation of a K-hole, the Auger cascade, and Coulomb explosion (ion dissociation). This sequence is captured in the general formula for the cross section of Auger defect formation

$$\sigma_d = \sigma_K \alpha_A \exp\left(-\frac{\tau_+}{\langle\tau_e\rangle}\right). \tag{1}$$

Here, $\sigma_K$ represents the cross section for K-ionization [15], $\alpha_A$ denotes the probability of forming a multiple Auger charge (Z) [3]. The expression $P = \exp(-\tau_+/\langle\tau_e\rangle)$ [9] indicates the probability of ion dissociation occurring in the time frame shorter than that required for electronic relaxation. Despite significant experimental evidence and theoretical models of Auger destruction in molecular and solid systems, the mechanism of destruction and defect formation remain incompletely understood, particularly regarding the final stage, the Coulomb explosion, which competes with the electronic neutralization process. In this study, we will illustrate the probabilistic origin of ion dissociation resulting from the Auger cascade, which contends with the quantum process of ion neutralization by valence electrons. We will first calculate the critical time for ion dissociation based on energy conservation, followed by the presentation of a probabilistic model for two parallel competitive processes: classical atomic interactions and quantum electronic processes.

The critical time for ion dissociation ($\tau_+$) required to accumulate sufficient kinetic energy to overcome binding energy can be expressed in terms of the potentials associated with the ground state of a neutral atom and its excited ionized state (see Fig. 1). As a result of the Auger cascade, the atom transits from its neutral ground state ($U_{gr}(R)$) to a multi-charged excited state ($U_{ex}(R)$), which is vertical, in accordance with the Franck-Condon principle. The excited ionic state is unstable and leads to ions separating from one another. The critical time for ion displacement can be determined using the following energy conservation law

$$U_{ex}(R) = U_{ex}(R_o) - \frac{\mu}{2}\left(\frac{dR}{dt}\right)^2. \tag{2}$$

Here, μ represents the reduced mass of the ions, and $R_o$ is the equilibrium distance between the atoms. From Eqn.2, we can derive the formula for $\tau_+$

$$\tau_+ = \sqrt{\frac{\mu}{2}} \int_{R_0}^{R_0+\Delta R} (U_{ex}(R_o) - U_{ex}(R))^{-1/2} dR = \sqrt{\frac{\mu}{2Ze^2}} \int_{R_o}^{R_o+\Delta R} \left(\frac{1}{R_o} - \frac{1}{R}\right)^{-1/2} dR \tag{3}$$

There is a critical distance (ΔR) from the equilibrium distance (R₀) at which the kinetic energy generated from the repulsive interaction between the ions is equal to the dissociation energy, given by

$$U_{gr}(\infty) - U_{gr}(R_o + \Delta R) = U_{ex}(R_o) - U_{ex}(R_o + \Delta R). \tag{4}$$

If the distance between the ions exceeds the critical value, the kinetic energy of the ions will surpass the binding energy, leading to the dissociation of atoms, even if the positively charged states are neutralized by electrons.

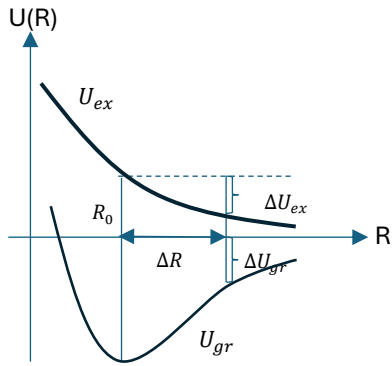

**Fig.1.** The kinetic energies of multiple ionized atoms separating from one another compete against the binding energy between the atoms

Thus, the Auger-destruction cross-section is inversely proportional $\tau_+$ which decreases with the charge as $1/\sqrt{Z}$. For DNA molecules, this implies that $\tau_+$ will be smallest for phosphorus compared to other elements, as phosphorus has more electron shells in its atomic structure.

Now, let's examine the process of ion displacement due to repulsion and the accumulation of kinetic energy occurring simultaneously with their neutralization by capturing electrons from the valence band. The probability of ions colliding with electrons within a time interval dt is $dt/\langle \tau_e \rangle$, where $\langle \tau_e \rangle$ represents the mean time for neutralization by electrons. We can define $P(t)$ as the probability of an ion not colliding with electrons up to time t. Thus, $P(t + dt)$ is the probability of the ion not colliding up to time t, multiplied by the probability of not colliding in the subsequent time interval dt, or in other words,

$$P(t + dt) = P(t)\left(1 - \frac{dt}{\langle \tau_e \rangle}\right) \tag{5}$$

From calculus we have
$$P(t+dt) = P(t) + \frac{dP(t)}{dt}dt \tag{6}$$
By equalizing these equations, we receive
$$\frac{1}{P}\frac{dP}{dt} = -\frac{1}{\langle \tau_e \rangle} \tag{7}$$
If the ion accumulates sufficient kinetic energy to break the chemical bond within the critical time $\tau_+$ without colliding with electrons, destruction occurs. Consequently, the integration should be performed from $t=0$ to $t=\tau_+$, resulting in the following solution
$$P(\tau_+) = e^{-\tau_+/\langle \tau_e \rangle} \tag{8}$$

In solid states, the condition for the occurrence of Coulomb explosion is that the ion dissociation time must be shorter than the instant neutralization time by electrons, specifically $\tau_+ < \tau_e(E)$ (Fig.2), where E represents the energy of electrons in the valence band.

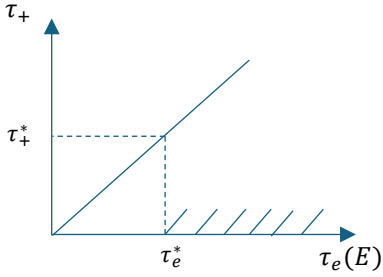

**Fig.2.** For a certain $\tau_+^*$, the dissociation occurs only if $\tau_e(E) > \tau_e^* = \tau_+^*$.

As mentioned earlier, $\tau_+$ is a function of Z and remains constant for a given Z. This allows us to outline a simple model of the neutralization process. For the instant neutralization time, we can write the following expression
$$\tau_e(E) = \frac{R}{v_g} = \frac{R}{\hbar^{-1}dE(k)/dk} \tag{9}$$
Here $R$ is the mean distance for electrons to reach the potential well of Z charged ions, $v_g$ is the group velocity of an electron as a wave packet. The effects of the crystal on the electron are contained in the dispersion relation, $E(k)$. In case of the effective mass-approach, $E(k) = \frac{\hbar^2 k^2}{2m_{eff}}$, where $m_{eff}$ is the effective mass of electron, we have
$$\tau_e(E) = \frac{Rm_{eff}}{\hbar k} \tag{10}$$

Thus, the larger the effective mass of the electron in the crystal, the longer the neutralization time. A mean neutralization time can be found from the following expression

$$\langle \tau_e \rangle = \int_0^{E_v^c} \tau_e(E) N(E)\, dE = R\sqrt{\frac{m_{eff}}{2}} \int_0^{E_v^c} N(E) \frac{dE}{\sqrt{E}} \tag{11}$$

Here $N(E)$ is the density of electronic states in the valence band, $E_v^c = \Delta E_v/2$ is the midpoint of the valence band and the reference point for energy is the bottom of the valence band. The integration should be performed from the bottom of the valence band to its midpoint, as the effective mass is negative in the upper portion of the band, causing the electrons to repel from the positive charge electric field. In case of very narrow valence band, $N(E) = N_0 \delta(E - E_c)$, we have

$$\langle \tau_e \rangle = R\sqrt{\frac{m_{eff}}{2}} \frac{N_o}{E_c^{1/2}} = \sqrt{\frac{m_{eff}}{2}} \frac{N_o}{(\Delta E_v)^{1/2}} \tag{12}$$

Thus, the mean neutralization time is inversely proportional to the width of the valence band, $\langle \tau_e \rangle \propto \frac{1}{(\Delta E_v)^{1/2}}$. Consequently, the wider valence band, the shorter mean neutralization time. This indicates that the destruction via the Auger cascade occurs in ionic crystals when the width of the valence bands meets a specific condition: $\tau_+ < \langle \tau_e \rangle = \frac{const}{(\Delta E_v)^{1/2}}$ (Fig.3). As noted earlier, Dexter reached a similar conclusion using the uncertainty principle, while we derived this relationship from the probabilistic nature of ion-electron collisions.

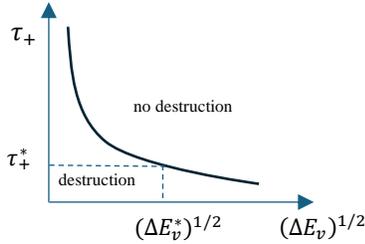

**Fig.3.** For a certain $\tau_+^*$, the destruction occurs if $(\Delta E_v)^{1/2} < (\Delta E_v^*)^{1/2} = const/\tau_+^*$.

By taking into account $\tau_+ \sim 1/\sqrt{Z}$, we can assume that when a certain $Z^*$ charge is formed by Auger-cascade, the Auger-destruction occurs in crystals with the width of valence band satisfying the condition: $\Delta E_v < \Delta E_v^* = const * Z^*$ (Fig.4).

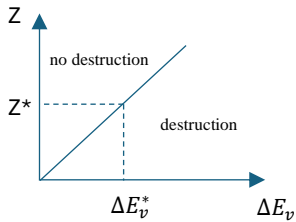

**Fig.4.** For a certain $Z^*$, the destruction occurs if $\Delta E_v < \Delta E_v^* = const * Z^*$

Thus, for unstable multi-charged ions generated in molecular and solid systems through Auger cascades, the destruction cross-section is directly proportional to the probability of Coulomb explosion determined by two characteristic times of atom dissociation and electron neutralization. It was shown that as the time for atomic dissociation, $\tau_+$, decreases with the charge according to the relationship $1/\sqrt{Z}$. The probability of the Coulomb explosion leading to the Auger-destruction has been obtained based on the kinetic theory of gases. By using approximation of the density of states in the valence band as a delta function, it was shown that the destruction via the Auger cascade occurs in ionic crystals when the width of the valence band meets the following specific condition: $\tau_+ < \frac{const}{(\Delta E_v)^{1/2}}$.